\documentclass[journal=jctcce,manuscript=article]{achemso}
\setkeys{acs}{articletitle=true,doi=false}
\usepackage[version=4]{mhchem} 
\usepackage{textcomp}
\usepackage{graphicx}
\usepackage{hyperref}
\usepackage{color}
\usepackage{soul}
\newcommand{\doi}[1]{\href{http://dx.doi.org/#1}{\nolinkurl{#1}}}
\newcommand{\rvec}{\mathbf{r}}

\author{Marco Cazzaniga}
\affiliation[SCITEC]{Consiglio Nazionale delle Ricerche, Istituto di Scienze e Tecnologie Chimiche (CNR-SCITEC), 20133 Milano, Italy}
\altaffiliation{Present address: Department of Chemistry, University of Milan, via Golgi 19, 20133 Milano, Italy}

\author{Fausto Cargnoni}
\email{fausto.cargnoni@cnr.it}
\affiliation[SCITEC]{Consiglio Nazionale delle Ricerche, Istituto di Scienze e Tecnologie Chimiche (CNR-SCITEC), 20133 Milano, Italy}

\author{Marta Penconi}
\affiliation[SCITEC]{Consiglio Nazionale delle Ricerche, Istituto di Scienze e Tecnologie Chimiche (CNR-SCITEC), 20133 Milano, Italy}

\author{Alberto Bossi}
\email{alberto.bossi@cnr.it}
\affiliation[SCITEC]{Consiglio Nazionale delle Ricerche, Istituto di Scienze e Tecnologie Chimiche (CNR-SCITEC), 20133 Milano, Italy}

\author{Davide Ceresoli}
\email{davide.ceresoli@cnr.it}
\affiliation[SCITEC]{Consiglio Nazionale delle Ricerche, Istituto di Scienze e Tecnologie Chimiche (CNR-SCITEC), 20133 Milano, Italy}

\title[GW-BSE on Ir(III) complexes]{Ab-initio Many Body Perturbation Theory
calculations of the electronic and optical properties of cyclometalated Ir(III) complexes}

\begin{document}

\begin{tocentry}
\includegraphics[width=9cm]{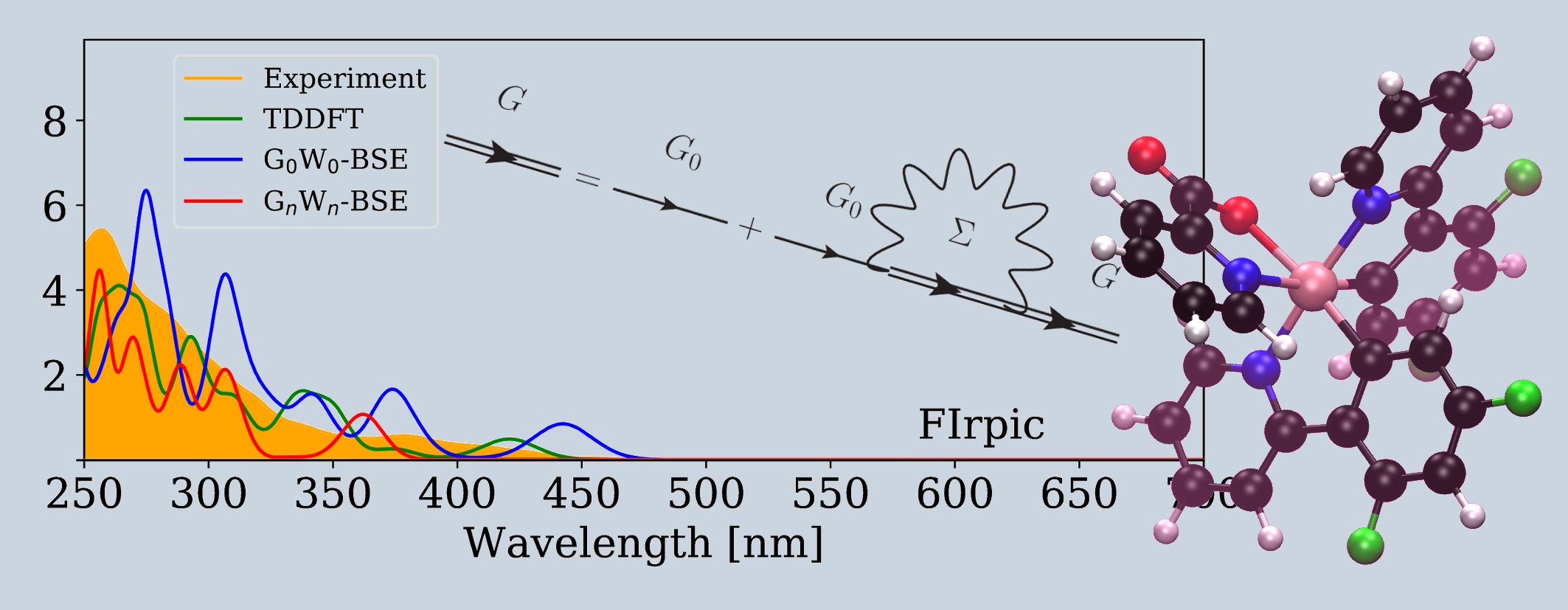}
\end{tocentry}

\begin{abstract}
Cyclometalled Ir(III) compounds are the preferred choice as organic emitters
in Organic Light Emitting Diodes. In practice, the presence of the transition
metals surrounded by carefully designed ligands allows the fine tuning of the
emission frequency as well as a good efficiency of the device.
To support the development of new compounds the experimental measurements
are generally compared with ab-initio calculation of the absorption
and emission spectra. The standard approach for these calculations is TDDFT
with hybrid exchange and correlation functional like the B3LYP. Due to
the size of these compounds the application of more complex quantum
chemistry approaches can be challenging. In this work we used Many Body
Perturbation Theory approaches (in particular the GW approximation
with the Bethe-Salpeter equation) implemented in gaussian basis sets,
to calculate the quasiparticle properties and the adsorption spectra of
six cyclometalled Ir(III) complexes going behind TDDFT.
In the presented results we compared standard TDDFT simulation with
BSE calculations performed on top on perturbative G$_0$W$_0$ and
accounting for eigenvalue self consistency. Moreover, in order
to investigate in detail the effect of the DFT starting point,
we concentrate on Ir(ppy)$_3$ performing GW-BSE simulations
starting from different DFT exchange and correlation potentials.
\end{abstract}

\section{Introduction}
Phosphorescent Organic Light Emitting Diodes (phOLEDs) are nowadays widely
employed in a large number of devices, like portable light sources and displays.
A huge research effort is devoted in designing new emitters with improved
efficiency and operational stability.~\cite{Minaev2014,Tao2014,Baranoff2015,Kesarkar2016,Penconi2017,Penconi2018}

Cyclometalated Ir(III) complexes are the preferred emitters molecules in
these devices, because of their large quantum efficiency approaching unity.
The large Spin Orbit Coupling (SOC) of Iridium is responsible for an efficient
phosphorescent emission and the low-lying electronic transitions display
a mixed Ligand-Center Metal-to-Ligand-Charge-Transfer (LC-MLCT) character.
By tailoring the chemical nature of the ligands, it is possible to tune
the emission frequency from the near infrared (NIR) to deep blue. Additionally,
replacing one ligand in homoleptic compounds with a different one
(giving rise to heteroleptic complexes) can improve the emitter performance,
increase the stability of the devices, and finely tune the emission frequency
in order to achieve the desired ``color''.~\cite{Gu2008,Li2013}

To design new emitters \emph{in-silico}, theory and computations
are required to understand and predict the measured photo-physical properties
of cyclometalated complexes. The standard theoretical approach for this
kind of investigation is Density Functional Theory (DFT) together with its
Time Dependent extension (TDDFT). Hybrid functionals (typically B3LYP, PBE0),
long-range corrected hybrids (i.e. LC-PBE~\cite{LCPBE}, LC$\omega$-PBE~\cite{LComegaPBE})
and range-separated hybrids (i.e. CAM-B3LYP~\cite{CAMB3LYP}) are the most
used to describe the optical absorption and emission spectra of these systems.
As it turns out, the ground state properties as well as the TDDFT-calculated
optical absorption and emission spectra, result in with a good agreement
with experiments.~\cite{Jacquemin2007,Asada2009}

Anyway, the aforementioned good performances of DFT/TDDFT with hybrid
functionals is somehow fortuitous. Going beyond this standard approach with
wavefunctions methods, is at present unfeasible because of the large number
of active electrons typical of these complexes. Accordingly, only a limited number
of authors tried to adopt different approaches to benchmark or improve
the accuracy of theoretical predictions. Examples are the application of
the DFT+Hubbard~U method~\cite{Himmetoglu2012}, the explicit inclusion of
SOC~\cite{Smith2011,Younker2013,Brahim2014,Mori2014} and the use of
multireference approaches.~\cite{Koseki2013,Heil2016}

Many Body Perturbation Theory (MBPT), particularly the GW and Bethe Salpeter
Equation (BSE)~\cite{Salpeter1951} methods provide an alternative theoretical framework to access
both ground and excited state properties of condensed matter~\cite{Hedin1965,Hybertsen1986,Onida2002,Martin2016,Reining2017,Golze2019}
and molecular systems.~\cite{Faber2014,Bruneval2016,Marom2017,Blase2018,Lettmann2019}
Most importantly, the BSE method appears to describe on an equal foot local
and charge-transfer excitations~\cite{Blase2011,Ziaei2016,Jacquemin2017a},
without fine tuning of the Hartee-Fock fraction or Coulomb range separation.

In this work, we apply the GW-BSE approachs to cyclometalated Ir-complexes
of technological relevance in the field of organic light emitting diodes (OLEDs).
These complexes are constituted of 52--61 atoms
and represent a computational challenge for the aforementioned approach, both
from the CPU and memory requests. Similalry to Refs.~\cite{Bruneval2013,Jacquemin2015a,vanSetten2015,Caruso2016,Maggio2017,Govoni2018,Gui2018},
in which the various levels of approximations of GW-BSE were benchmarked on selected sets
of small organic molecules, we compare our GW results obtained with different
level of self-consistency and with different DFT starting points.

We consider six Ir-complexes, with absorption and emission wavelengths spanning
across the entire visible spectrum, and we calculate the optical absorption spectra
starting from the B3LYP orbitals, using both the perturbative G$_0$W$_0$ and the
eigenvalue self-consistent G$_n$W$_n$ method. Then we discuss how discuss how
the quasiparticle energies and the optical absorption of Ir(ppy)$_3$, chosen
as reference complex, are affected by the DFT starting point.
In particular, we investigate how the results change when considering a pure local
functional (BLYP), two standard hybrids with different percentage of exact exchange
(B3LYP and BHLYP), and a Coulomb Attenuated functional (CAM-B3LYP).
The calculated optical absorption spectra are compared to experimental results
obtained in our group, reported in previous papers~\cite{Penconi2018,Penconi2019}
and on newly synthesized complexes.

\section{Theory and methods}\label{sec:theory_and_methods}
\subsection{Review of the theory}\label{sec:theory}
The theoretical approaches adopted for the computations discussed in the present work are based on 
the Many Body Perturbation Theory~\cite{Mahan1981,Fetter2003,Martin2016}. Within this framework the single quasiparticle properties can be accessed through the Hedin's equations, which consist in the following set of five equations~\cite{Hedin1965}:
\begin{eqnarray}
\label{eq:1hedin}
G(1,2)&=&G_0(1,2)+\int G_0(1,3)\Sigma(3,4)G(4,2)\,d3d4\\
\label{eq:2hedin}
\Gamma(1,2;3)&=&\delta(1,2)\delta(1,3)+\int\frac{\delta\Sigma(1,2)}{{\delta}G(4,5)}G(4,6)G(7,5)\Gamma(6,7;3)d4d5d6d7\\ 
\label{eq:3hedin}
\chi(1,2)&=&-i\int G(1,3)G(4,1)\Gamma(3,4;2)d3d4\\
\label{eq:4hedin}
W(1,2)&=&v_C(1,2)+\int v_C(1,3)\chi(3,4)W(4,2)d3d4\\
\label{eq:5hedin}
\Sigma(1,2)&=&i\int G(1,3)W(4,1)\Gamma(3,2;4)d3d4,
\end{eqnarray} 
where $G$ is the interacting Green function, $G_0$ the non-interacting one, $v_C$ is the bare Coulomb potential, $W$ is the screened interaction; $\chi$ is the polarizability, $\Sigma$ is the self-energy, 
and $\Gamma$ is the vertex function.
We adopt the notation in which a coordinate as "$1$" stands for the set of position, time and spin variables $({\bf r}_1,t_1,\sigma_1)$.

A self-consistent solution of the set of equations \eqref{eq:1hedin}-\eqref{eq:5hedin} is a challenging task and a series of approximations is applied to reduce the complexity of the problem.
The most relevant is to neglet the vertex function which yields to the so called GW approximation from the aspect of Eq. \ref{eq:5hedin} upon the aforementioned assumption.

Despite the neglect of the vertex function, a self-consistent solution of the GW equations remains challenging and computationally demanding. The starting point of a large part of these simplified approaches is the quasiparticle equation:
\begin{eqnarray}\label{qp:eq}
\left(-\frac{\nabla^2}{2}+V_{ext}+V_{H}\right)\Psi_{i}({\bf r})+\int d{\bf r}' \Sigma({\bf r},{\bf r}',E_{i}) \Psi_{i}({\bf r}') = E_{i} \Psi_{i}({\bf r})
\end{eqnarray} 
It should be noticed that Eq. \ref{qp:eq} is a Schr\"odinger like equation where the self energy $\Sigma$ is a complex and non Hermitian potential.

From the first applications of the GW approximation to bulk silicon\cite{Hybertsen1985,Hybertsen1986,Godby1988} a further simplified scheme has taken root. It consists in performing only a single iteration of Hedin's equations, thus obtaining the quasiparticle energies as a perturbative correction of the DFT-KS eigenvalues (we will refer to this approach as G$_0$W$_0$):   
\begin{eqnarray}\label{eig:eq}
E_i\simeq\epsilon_i + \langle \psi_{i}|\Sigma(E_i) - V_{xc}|\psi_{i} \rangle,
\end{eqnarray} 
where the solution for $E_i$ can be found either by linearizing the frequency dependency of the self-energy around $\epsilon_i$ or, as for the present case, by zero finding algorithms. 
This approach revealed successful in providing good band gaps in bulk materials but presented the drawback of retaining a strong dependency on the underlying approximations used in determining $G_0$ (the V$_{xc}$ functional of the DFT).

This is one of the arguments that motivate to go behind a mere perturbative calculation, toward self consistent GW. Anyway, even if some results are available, simplified approaches to self consistency have been proposed. In particular Faleev and coworkers introduced the so called Quasiparticle Self-Consistent GW (qsGW) \cite{Faleev2004,Kotani2007}. 
The basic idea is to design a static and Hermitian approximation for the self-energy, whose results are close to the full GW. Within the qsGW approach, the self-energy assumes the following expression:
\begin{eqnarray}\label{qpgw_sigma:eq}
\left\langle \Psi_{i}\left| \Sigma^{qsGW} \right| \Psi_{j} \right\rangle = \frac{1}{2}\, \mathrm{Re} \left\langle \Psi_{i}\left| \Sigma(\epsilon_i)+\Sigma(\epsilon_j) \right| \Psi_{j} \right\rangle
\end{eqnarray}
A qsGW calculations deals with successive evaluations of the matrix elements of Eq. \ref{qpgw_sigma:eq}, followed by a diagonalization of the quasiparticle equation until self-consistency is reached.  
A further simplification of this approach consists in updating only the quasiparticle eigenvalues while keeping fixed the starting point orbitals (we will refer hereafter to this approach as G$_n$W$_n$). 

Within MBPT neutral excitation energies can be accessed through the Bethe-Salpeter equation \cite{Salpeter1951}, which consists in the following Dyson like equation for the two-particle correlation function L:
\begin{eqnarray}\label{bse:eq}
\nonumber
L(1,2,3,4)\!\!\!&=&L_0(1,2,3,4)+\\
&+&\!\!\!\!\int L_0(1,2,5,6)\left[ v_C(5,7)\delta(5,6)\delta(7,8) + \frac{\delta\Sigma(5,6)}{\delta G(7,8)}\right]L(7,8,3,4)\, d5d6d7d8
\end{eqnarray}
where $L_0$ is the non-interacting correlation function. 
Analogously to the case of Hedin's equations, in practical applications of Eq.~\ref{bse:eq} some simplifying assumptions are made: $L_0$ is build from the GW quasiparticle energies and the corresponding orbitals, while the kernel in Eq. \ref{bse:eq} is obtained assuming a GW self-energy in its static ($\omega\rightarrow0$) limit and neglecting the $\delta W/\delta G$ term. 
Thanks to these approximations it is possible to rewrite the BSE as an eigenvalue problem in the particle-hole space, similar to the Casida's formalism for the TDDFT~\cite{Casida1995}:
\begin{eqnarray}\label{bse_matrix:eq}
\begin{pmatrix}
 A  & B \\ -B^* & -A^*
\end{pmatrix}
\begin{pmatrix}
X \\ Y
\end{pmatrix}
=
\omega
\begin{pmatrix}
X \\ Y
\end{pmatrix}
\end{eqnarray}
where the matrix A and B are respectively:

\begin{eqnarray}\label{bse_ab:eq}
\nonumber
A_{ia,jb}&=& (E_a-E_i)\delta_{ij}\delta_{ab} 
   -\alpha^{S/T} \left(ia |v_C| jb\right) + \left(ij|W|ab\right) \\
B_{ia,jb}&=& -\alpha^{S/T}\left(ia|v_C|bj\right) + \left(ib|W|aj\right)
\end{eqnarray}
where $i,j$ are occupied state and $a,b$ are virtual ones; $\alpha^{S/T}$ is 2 for singlet final
states and 0 for triplet ones; $(ij|V|kl) = \int d\rvec \int d\rvec' \psi_i(\rvec)^\star\psi_j(\rvec) V(\rvec,\rvec') \psi_k(\rvec')^\star\psi_l(\rvec')$
is a two-electron integral, and $W(\rvec,\rvec') = W(\rvec,\rvec',\omega\rightarrow0)$.
The diagonalization of this matrices allows to obtain the excitation energies and oscillator strengths, which are the necessary ingredients to simulate optical spectroscopies.

\subsection{Basis set convergence}
\begin{figure}[h!]
\centering
\begin{tabular}{cccc}
\includegraphics[scale=0.3]{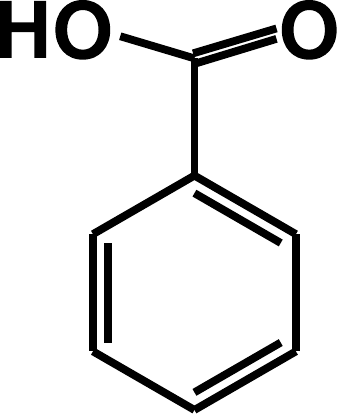} & \includegraphics[scale=0.3]{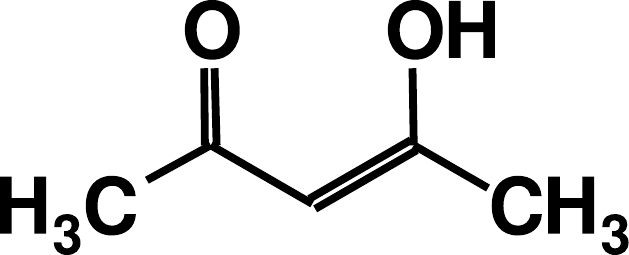} &
\includegraphics[scale=0.3]{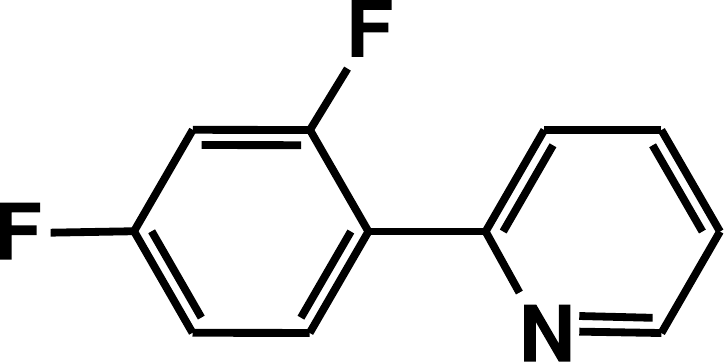} & \includegraphics[scale=0.3]{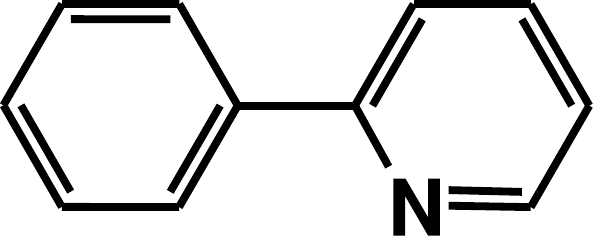} \\
pic & acac & dfppy & ppy\\
\includegraphics[scale=0.3]{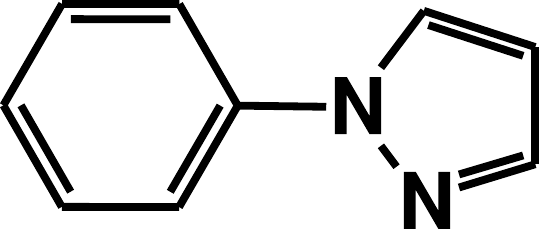} & \includegraphics[scale=0.3]{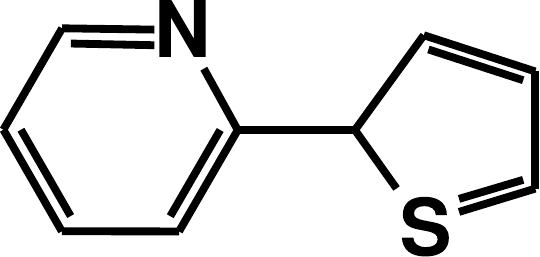} &
\includegraphics[scale=0.3]{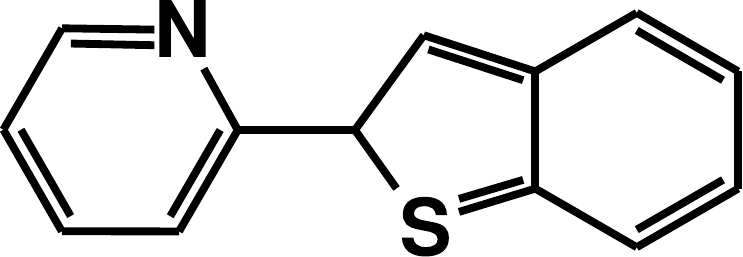} & \includegraphics[scale=0.3]{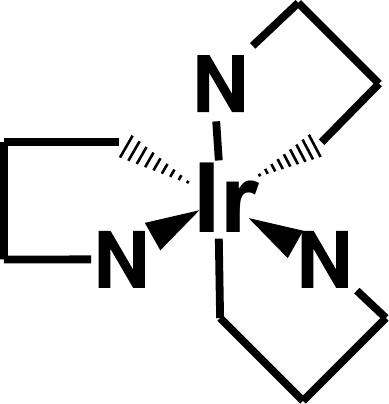} \\
ppz & thpy & btpy & Ir-small\\
\end{tabular}
\caption{Set of molecules used to analyze the basis set convergence of G$_0$W$_0$ and BSE
calculations. For the \emph{acac} molecule we choose the enolic form, which is more stable
than the keto form in gas phase. The \emph{Ir-small} complex is taken from Ref.~\cite{Koseki2013}}
\label{fig:frag}
\end{figure}

\begin{figure}[h!]
\centering
\includegraphics[width=0.8\columnwidth]{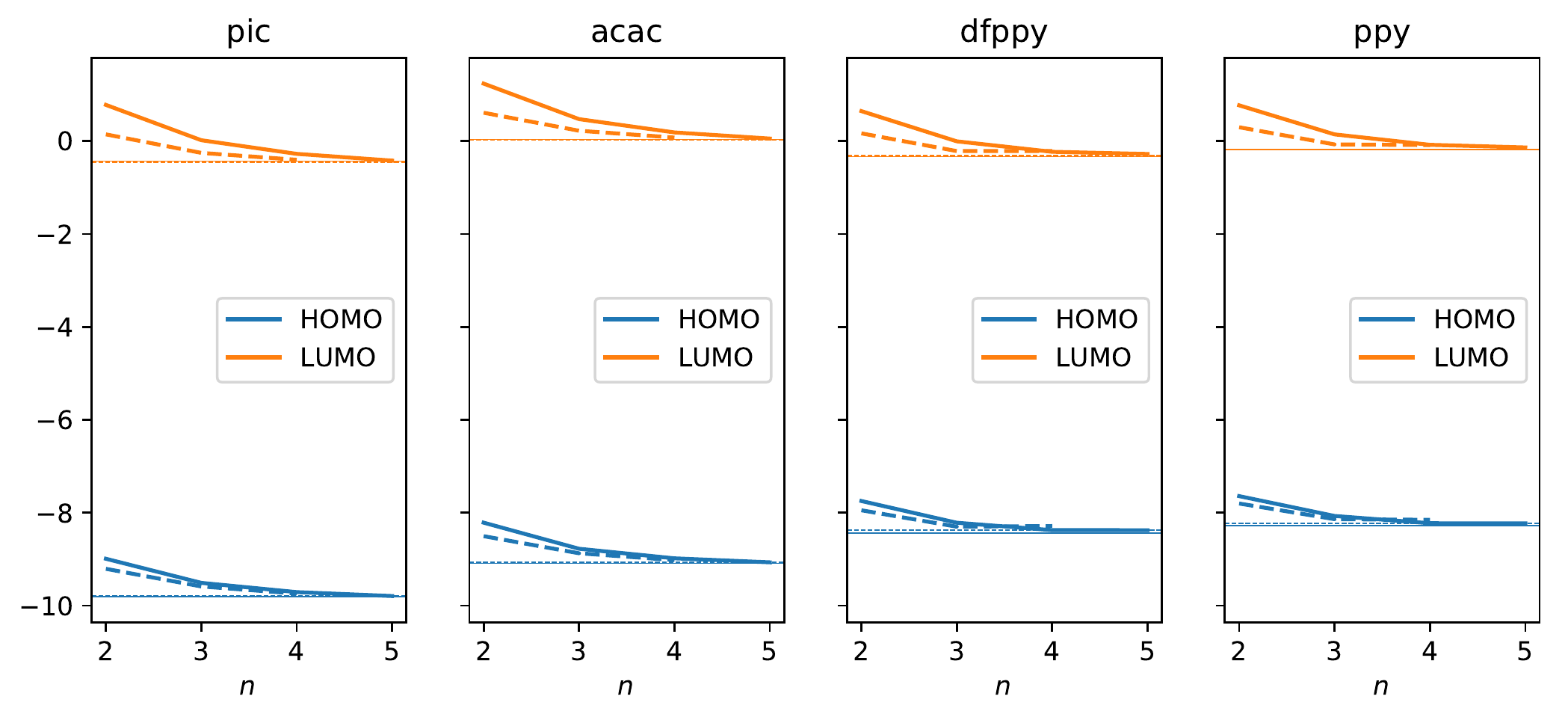}\\
\includegraphics[width=0.8\columnwidth]{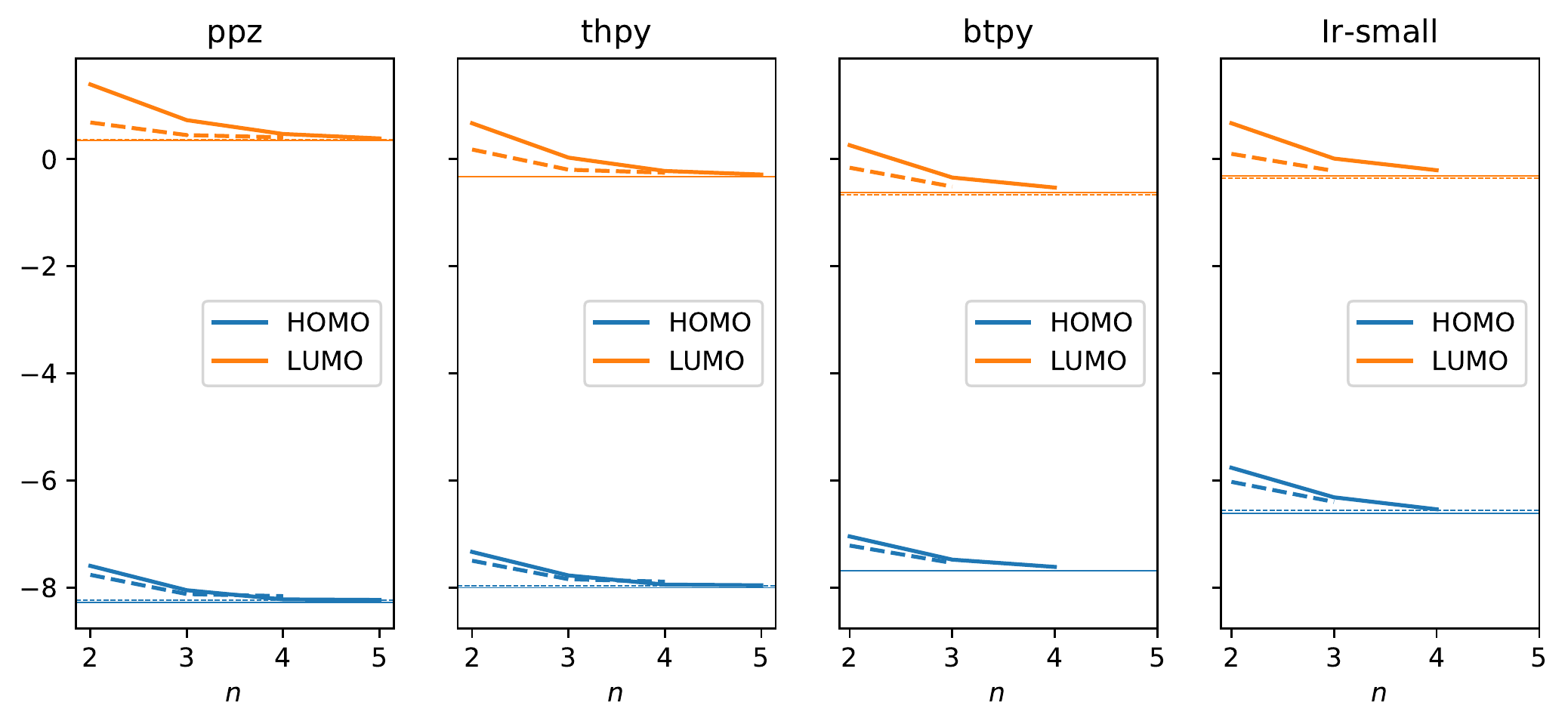}\\
\caption{Calculated G$_0$W$_0$ HOMO and LUMO as a function of the basis set size. Solid thick
lines: cc-pV$n$Z. Dashed thick lines: aug-cc-pV$n$Z. Thin lines: complete basis set extrapolation
as $E_\infty+A/n^3$.}\label{fig:GWfrag}
\end{figure}

\begin{figure}[h!]
\centering
\includegraphics[width=0.8\columnwidth]{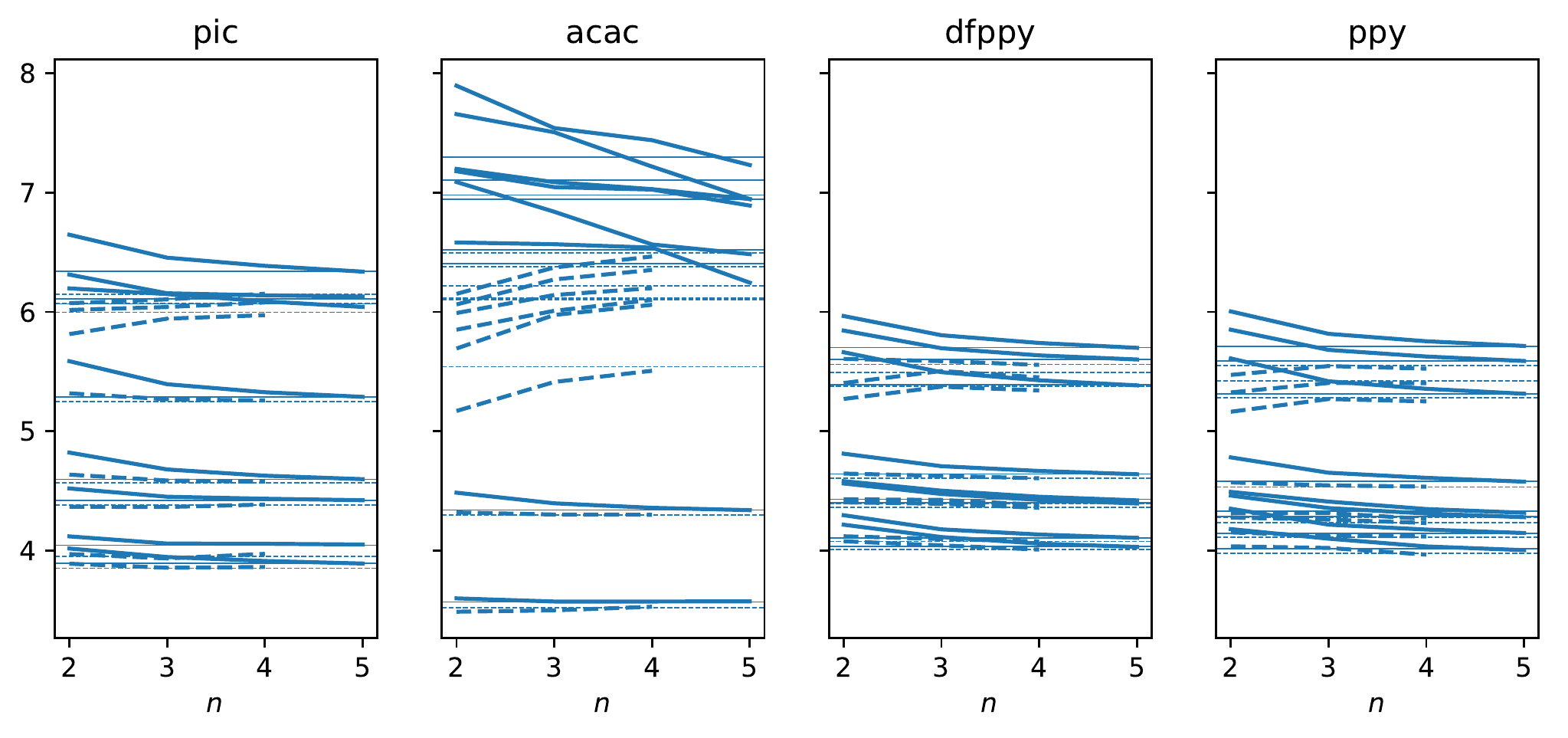}\\
\includegraphics[width=0.8\columnwidth]{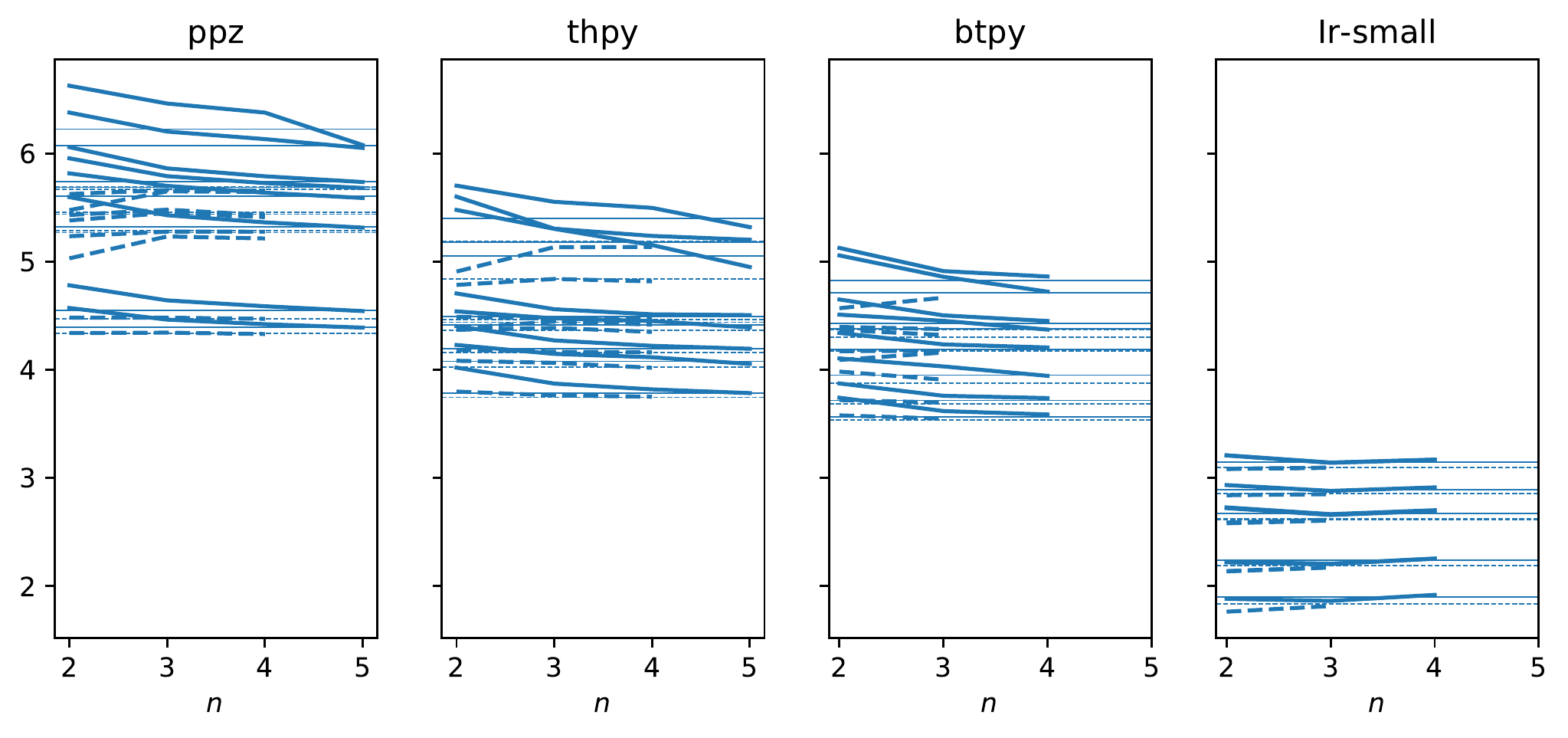}\\
\caption{Calculated low lying BSE excitation energies as a function of the basis set size. Solid thick
lines: cc-pV$n$Z. Dashed thick lines: aug-cc-pV$n$Z. Thin lines: complete basis set extrapolation
as $E_\infty+A/n^3$.}\label{fig:BSEfrag}
\end{figure}

The basis set convergence of GW and BSE calculations has been studied and reported
extensively in Refs.~\cite{Bruneval2013,Bruneval2016,Jacquemin2016a,Rangel2016,Gui2018}.
We first analyzed the basis set convergence of both G$_0$W$_0$ and BSE calculations, on a small
set of eight organic molecules that constitute the cyclometalated complexes under study,
using the B3LYP functional.
The molecules are reported in Fig.~\ref{fig:frag}. We used the Dunning basis set series cc-pV$n$Z and
aug-cc-pV$n$Z. The G$_0$W$_0$ HOMO and LUMO are reported in Fig.~\ref{fig:GWfrag}.
As shown in Fig.~\ref{fig:GWfrag}, both the HOMO and LUMO converge from above as a function
of the basis set size. For this set of molecules, the augmented basis set converges faster.
Our results show that the cc-pVTZ basis set is capable to predict HOMO and LUMO energies within $\sim$0.1~eV
of the largest basis set.
Likewise, we report in Fig.~\ref{fig:BSEfrag} the first eight low lying excited states computed
with BSE. Our results show again that the cc-pVTZ basis set is capable to predict the first
excitations energies within $\sim$0.1~eV of the most converged basis set. One notable exception
is \emph{acac} where only the first two low lying excitations are well converged with the cc-pVTZ.
In this case, there are two excited states whose energy decreases steeply with the basis set size,
and crosses with other well-converged excited states. This situation is similar to what is reported
in Ref.~\cite{Bruneval2015} for the pyrrole molecule. We note in passing that benzene, pyrrole,
furane and thiopehene display a problematic convergence with respect to the basis set (see Fig.~S26
of the supplementary information). However, our set of molecules shows a fast convergence
behavior with respect to basis set size.

Given our results and observations, we decided to employ the cc-pVTZ basis set in the subsequent
calculations, as it provides a convenient trade off between accuracy and computational cost. 

\label{sec:basis}

\subsection{Computational details}
\label{sec:computational}
\begin{figure}[htbp]
\begin{center}
  \includegraphics[width=8cm]{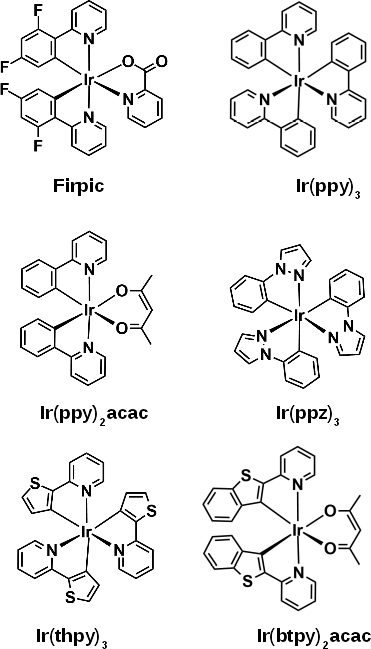}
  \caption{Scheme of the 6 complexes addressed in this work.}
  \label{fgr:molecules}
  \end{center}
\end{figure}

Next we performed excited state simulations on six different Ir(III) cyclometalled complexes, 
both homoleptic and heteroleptic, which represent the standard red, green or blue OLED emitters. 
In particular we addressed: FIrpic, Ir(ppy)$_3$, Ir(ppy)$_2$acac, Ir(ppz)$_3$, Ir(thpy)$_3$, and 
Ir(btpy)$_2$acac, whose molecular structures are represented in Fig. \ref{fgr:molecules}.
In the case of homoleptic compounds we considered the facial isomer, while for the heteroleptic complexes we selected the 
trans one, since the synthesis protocols generally produce these isomers.

Optimized geometries of the six complexes in their electronic ground state have been determined with the Gaussian09 package \cite{gaussian09}, adopting a 6-31G(d,p)
basis set for the light atoms and an aug-cc-pVDZ + ECP (Effective Core Potential) for Ir. 
Spin orbit coupling is not treated explicitly in our calculations, but is included through the ECP, at 
the scalar-relativistic level.  Smith and coworkers~\cite{Smith2011} showed that at the TDDFT level, SO has a negligible effect on absorption spectrum, but a large impact on the magnetic circular dichroism (MCD) spectra of
cyclometalated Ir(III) complexes.
All computations have been carried out at the B3LYP level of theory. In the case of Ir(ppy)$_3$, we tested the performance of three more functionals, namely BLYP, BHLYP and CAM-B3LYP.

Excited state calculations were performed with the MolGW (version 2.A) code
\cite{Bruneval2016,MolGW} at the TDDFT, GW and BSE levels of theory,
with the single exception of the TDDFT simulation of Ir(ppy)$_3$ with
the CAM-B3LYP $f_{xc}$, performed with Gaussian09.
We selected the \mbox{cc-pVTZ} basis set for all the atoms, and we adopted
the Resolution of Identity (RI) for the four center integrals.
Also in this case we used an ECP for describing the Ir core electrons.
In all computations we adopted the frozen core approximation.
We couldn't use the aug-cc-pVTZ for the large Ir(III) complexes because it
required too much memory. As we showed in Sec.~\ref{sec:basis}, the cc-pVTZ basis
set can provide converged results within $\sim$0.1~eV.

At TDDFT and BSE level we included only the lowest 500 orbitals (i.e. up
to 32--40~eV depending on the molecule),
which proved to ensure converged results. The eigenvalue self-consistent G$_n$W$_n$
calculations consisted in five iterations, and we observed that after the third
iteration, eigenvalues reached self-consistency to $10^{-3}$~eV.
We did not simulate vibrational side-bands, neither solvent effects.
We broadened the absorption spectra with a 0.1~eV wide gaussian function, which
yields more peaks/features than the experimental spectra.

The most expensive part of the calculation is the GW iteration. We run our
calculations on 8 cluster nodes, equipped with 2$\times$18-core Intel Xeon E5-2697v4 CPUs
and 128~Gb RAM per node. A full G$_n$W$_n$ (five iterations) calculation
required nearly 24 hours on this machines, and the peak memory usage reported by
the code was 1.3~Gb per MPI process (288 MPI processes). Conversely, each
BSE calculation required $\sim$1~hour with a peak memory usage of 0.4~Gb per MPI
process.

\section{Experiments}
\label{sec:measurements}
Of the six cyclometalated complexes FIrpic was purchased from
SigmaAldrich and used as received. The other complexes were synthesized following
the procedure reported in the following papers: Ir(ppy)$_3$~\cite{Dedeian1991};
Ir(ppy)$_2$acac~\cite{Lamansky2001}; Ir(ppz)$_3$~\cite{Tamayo2003}.
Ir(thpy)$_3$ was instead synthesized according to slight modified procedure
of Ref.\cite{McGee2007} using AgOTf instead of AgPF$_6$, starting from
[Ir(ThPy)$_2$Cl]$_2$ (21.8\% yield).
The absorption spectra of FIrpic, Ir(ppy)$_3$ and Ir(ppy)$_2$acac where
measuread and reported in previous works~\cite{Penconi2018,Penconi2019},
whereas, for the other complexes, UV/Vis absorption spectra were obtained
on Agilent 8453 spectrophotometer in 1~cm path length quartz cell with
dichloromethane.

\section{Quasiparticle and optical absorption of Ir(III) complexes}
\label{sec:complex}
\subsection{Quasiparticle energies}
\label{sec:complex_qp}
In Table \ref{tab:gapshomolumo} we reported the values of the HOMO-LUMO
gaps of the complexes addressed in the present work. At the DFT-B3LYP
level of theory the results are consistent with previously reported values.
In particular, ordering the complexes by increasing gaps we obtain the following sequence:
Ir(btpy)$_2$acac $<$ Ir(ppy)$_2$acac $<$ Ir(thpy)$_3$ $<$ Ir(ppy)$_3$ $<$ FIrpic $<$ Ir(ppz)$_3$.
Moving to the G$_0$W$_0$ corrected gaps we observe a large opening of the HOMO-LUMO gaps, as expected, which is further 
enlarged when the self-consistency on eigenvalues is included (G$_n$W$_n$).
Within these latter approaches we register a slight modification of the ordering of energy gaps,
which becomes:
Ir(btpy)$_2$acac $<$ Ir(ppy)$_3$ $<$ Ir(ppy)$_2$acac $<$ Ir(thpy)$_3$ $<$ FIrpic $<$ Ir(ppz)$_3$.
However, we note that at these levels of theory the gaps of Ir(ppy)$_3$, Ir(ppy)$_2$acac and Ir(thpy)$_3$
are almost equal.

Inspecting the frontier molecular orbitals reported in Table S1--S2 of the Supplementary Information, it 
is possible to observe that, in agreement with the expectations for this class of compounds, the 
highest occupied orbitals present significant contribution coming from the central Ir, while the lowest 
virtual orbitals are mainly located on the ligands.
In homoleptic compounds the frontier orbitals are spread
over all ligands, while this is not the case in heteroleptic complexes. Indeed, in FIrpic no contribution of the HOMO is found on the ancillary ligand, while the LUMO behaves at the opposite, being mostly located on the picolinate fragment.
In Ir(ppy)$_2$acac and Ir(btpy)$_2$acac, neither the HOMO nor the LUMO present relevant contributions on the acac moiety.
Even though we didn't impose symmetry, Ir(ppy)$_3$, Ir(ppz)$_3$ and Ir(thpy)$_3$ have C$_3$ symmetry
whose irreps can be singly (A) or doubly degenerate (E). The presence of nearly degenerate
eigenvalues is visible in Table~S3. Ir(ppy)$_2$acac and Ir(btpy)$_2$acac display instead C$_2$ symmetry
while FIrpic is C$_1$. Therefore, eigenvalue degeneracies can only be accidental in
these compounds.

\begin{figure}[htbp]
\begin{center}
  \includegraphics[width=\textwidth]{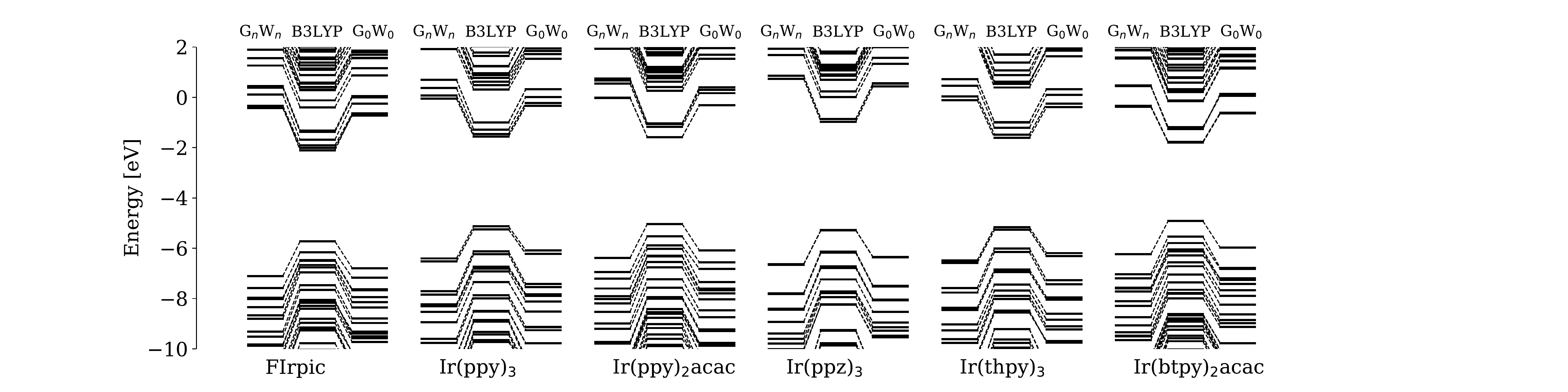}
  \caption{Quasiparticle energies of the six complexes addressed in the present work. 
  Comparison of eigenvalues obtained from a DFT-B3LYP calculations compared to G$_0$W$_0$ and G$_n$W$_n$ 
  results obtained on top of the B3LYP ground state.}
  \label{fgr:complexesquasiparticle}
  \end{center}
\end{figure}

In Fig. \ref{fgr:complexesquasiparticle} the trend of the quasiparticle energies calculated
at the DFT-B3LYP level are compared with the results obtained from the G$_0$W$_0$ and G$_n$W$_n$ approximations.
It is possible to notice how the opening of the HOMO-LUMO gap is coming from a decrease of the energies of
occupied orbitals and a comparable increase in the energy of the virtual orbitals.
Furthermore, with the exception of FIrpic, in the present series of compounds, neither the
G$_0$W$_0$ nor the G$_n$W$_n$ approximation modify the orbital ordering. In fact, as shown in Table~S3,
the quasiparticle correction to the B3LYP eigenvalues of FIrpic lead to an inversion between the LUMO
and LUMO+1. As we will see later, the inversion of quasiparticle orbitals will yield the
inversion in the orbitals involved in the first singlet excitation of FIrpic.

\begin{table}
  \caption{\label{tab:gapshomolumo}HOMO-LUMO gap (eV) of molecules addressed in the present work 
  calculated as difference of DFT-KS eigenvalues compared to G$_0$W$_0$ and G$_n$W$_n$ results.
  Calculations have been performed on top on a B3LYP DFT starting point.}
\begin{tabular}{|ll|lll|}
\hline
                  &                      & DFT-B3LYP & G$_0$W$_0$@B3LYP & G$_n$W$_n$@B3LYP \\ 
\hline
 FIrpic           & Present              & 3.614 & 6.066 & 6.673 \\
                  & Ref. \cite{Li2015}   & 3.76     &          &          \\
                  & Ref. \cite{Li2013}   & 3.73     &          &          \\
                  & Ref. \cite{Park2006} & 3.746    &          &          \\                  
 Ir(ppy)$_3$      & Present              & 3.569 & 5.745 & 6.354 \\
                  & Ref. \cite{Hay2002}  & 3.565    &          &          \\
                  & Ref. \cite{Park2006} & 3.660    &          &          \\                                    
 Ir(ppy)$_2$acac  & Present              & 3.449 & 5.763 & 6.361 \\
                  & Ref. \cite{Hay2002}  & 3.456    &          &          \\                    
 Ir(ppz)$_3$      & Present              & 4.305 & 6.783 & 7.374 \\
                  & Ref. \cite{Park2005} & 4.471    &          &          \\                    
 Ir(thpy)$_3$     & Present              & 3.546 & 5.800 & 6.372 \\
 Ir(btpy)$_2$acac & Present              & 3.119 & 5.334 & 5.857 \\
\hline
\end{tabular}
\end{table}

Before moving to the discussion of excited state properties, we compute and report
in Table~\ref{tab:gapsipea} the Ionization Potential (IP) and Electron Affinities (EA) of
the compounds, comparing the results with available experimental data.
As already shown by Korbel et al.~\cite{Korbel2014}, we notice
that the IP differences between the $\Delta$SCF and the GW results are rather small.
In detail the G$_0$W$_0$ values are slightly smaller than the $\Delta$SCF ones,
while the G$_n$W$_n$ are larger. Anyway, the three approaches provide values which
overestimate the available data.

The behavior of the EA is more complex. There is not a straightforward agreement
of $\Delta$SCF results against GW. This can be abscribed to the artificial stabilization of anions
in DFT using a localized basis set. Unfortunately, the exiguity of available experimental
data does not allow us to draw a definite conclusion about the performance of the methods,
regarding the determination the EA. Note that Ir(ppz)$_3$ is predicted to
have negative electron affinity, i.e. the anion is unstable with respect to the
neutral molecule and one electron at infinity. From the experimental point of
view, the Ir(ppz)$_3$ anion would exist as a long-lived resonance with the states
of the continuum.

\begin{table}
  \caption{\label{tab:gapsipea} Vertical ionization potential and electron affinity (eV) of the molecules 
  addressed in the present work calculated with the $\Delta$SCF approach compared to G$_0$W$_0$ and 
  G$_n$W$_n$ results.
  Calculations have been performed on top on a B3LYP starting point.}
\begin{tabular}{|l|ll|}
\hline
 & I.P. & E.A.  \\ 
\hline
 & \multicolumn{2}{c|}{FIrpic} \\ 
 Experiment \cite{DAndrade2005} & 5.91 & \\ 
 $\Delta$SCF Ref.\cite{Li2013} & 6.68 & 0.52 \\
 $\Delta$SCF Ref.\cite{Li2015} & 6.69 & 0.50 \\
 $\Delta$SCF & 6.886 & 0.900 \\ 
 G$_0$W$_0$ & 6.792 & 0.662 \\ 
 G$_n$W$_n$ & 7.102 & 0.429 \\ 
\hline
 & \multicolumn{2}{c|}{Ir(ppy)$_3$fac} \\ 
 Experiment \cite{Yoshida2015} & 5.27 & 1.86\\
 Experiment \cite{DAndrade2005} & 5.10 & \\ 
 G$_0$W$_0$ Ref.\cite{Hay2002} & 5.94 & 0.08 \\
 $\Delta$SCF & 6.187 & 0.451 \\ 
 G$_0$W$_0$ & 6.085 & 0.340 \\ 
 G$_n$W$_n$ & 6.403 & 0.050 \\ 
\hline
 & \multicolumn{2}{c|}{Ir(ppy)$_2$acac} \\
 G$_0$W$_0$ Ref.\cite{Hay2002} & 5.97 & 0.03 \\
 $\Delta$SCF & 6.197 & 0.383 \\ 
 G$_0$W$_0$ & 6.082 & 0.313 \\ 
 G$_n$W$_n$ & 6.385 & 0.011 \\ 
\hline
 & \multicolumn{2}{c|}{Ir(ppz)$_3$fac} \\ 
 Experiment \cite{DAndrade2005} & 5.03 & \\ 
 $\Delta$SCF & 6.424 & -0.181 \\ 
 G$_0$W$_0$ & 6.347 & -0.436 \\ 
 G$_n$W$_n$ & 6.655 & -0.739 \\ 
\hline
 & \multicolumn{2}{c|}{Ir(thpy)$_3$fac} \\ 
 $\Delta$SCF & 6.262 & 0.498 \\ 
 G$_0$W$_0$ & 6.191 & 0.391 \\ 
 G$_n$W$_n$ & 6.480 & 0.108 \\ 
\hline
 & \multicolumn{2}{c|}{Ir(btpy)$_2$acac} \\ 
 $\Delta$SCF & 6.031 & 0.682 \\ 
 G$_0$W$_0$ & 5.976 & 0.642 \\ 
 G$_n$W$_n$ & 6.233 & 0.376 \\ 
\hline
\end{tabular}
\end{table}

\subsection{Optical absorption}
\label{sec:complex_bse}

\begin{figure}[htbp]
\begin{center}
  \includegraphics[width=\textwidth]{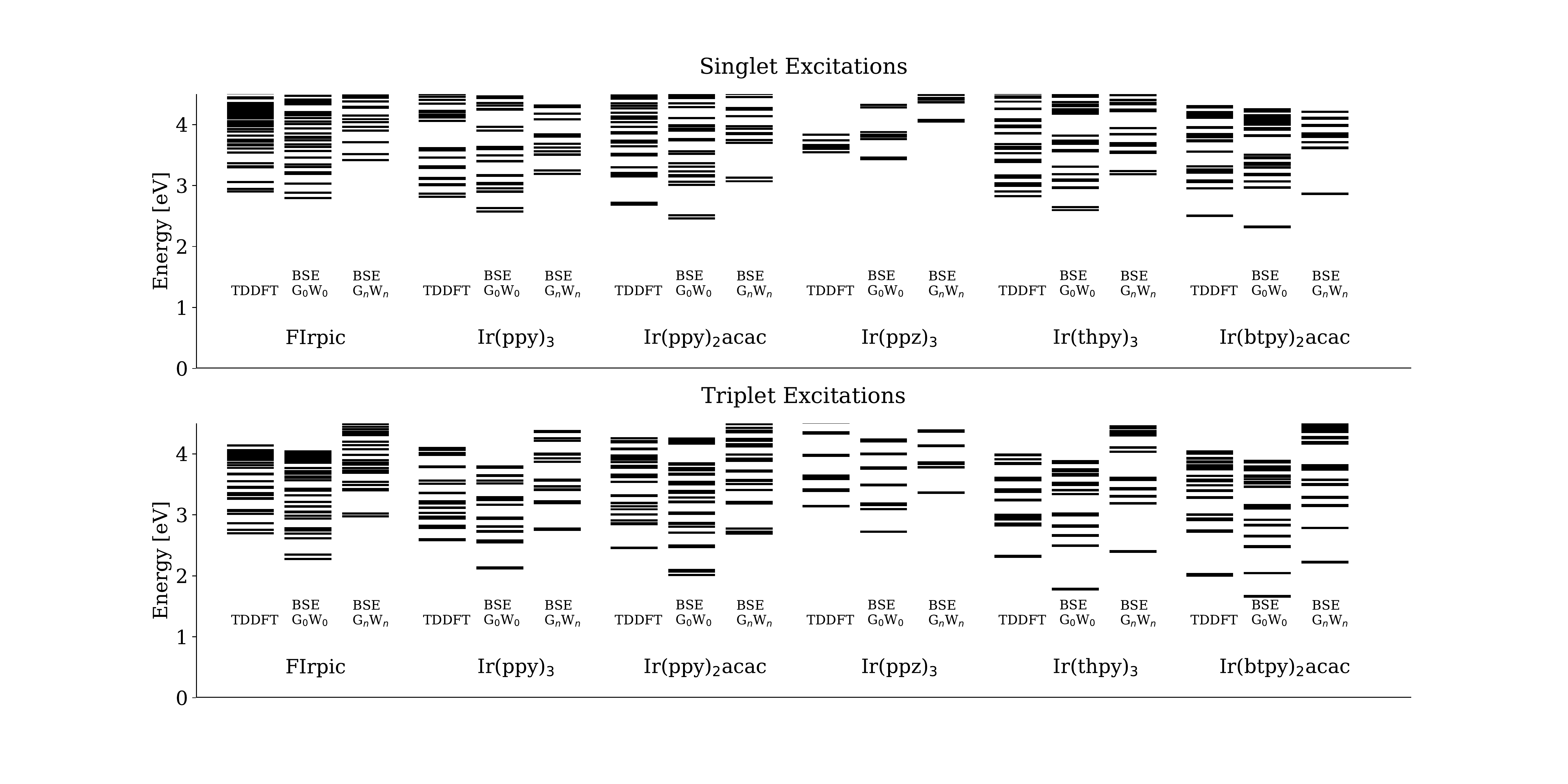}
  \caption{Lowest energy singlet and triplet excitation energies for the six complexes addressed in the 
  present work.
  Calculations have been performed at the TDDFT-B3LYP level as well as from the BSE on top on G$_0$W$_0$ 
  and G$_n$W$_n$ results.}
  \label{fgr:excitationsbse}
  \end{center}
\end{figure}

Figure \ref{fgr:excitationsbse} reports a plot of the excitation energies of the studied complexes,
which are collected in Table~S3--S15 of the supporting information.
The general trend for both singlet and triplet excitations compared to TDDFT results is a decrease of the
energies when a perturbative G$_0$W$_0$ is chosen as starting point for the BSE calculation, while we 
observe the opposite behavior for results obtained on top of a G$_n$W$_n$ calculation. Differences between BSE@G$_0$W$_0$ and 
BSE@G$_n$W$_n$ estimates of electron excitation energies are quite large, amounting to about 0.5 eV.

In more detail, in FIrpic we can observe that the four lowest singlets present essentially a single transition character. They involves transitions from HOMO-2, HOMO-1 and HOMO toward 
the orbitals from LUMO to LUMO+2, and present a mixed Metal to Ligand Charge Transfer (MLCT) and Ligand 
Centered (LC) character typical of these complexes. 
Comparing TDDFT and BSE results the most interesting difference is that the HOMO-LUMO excitation (which 
involves the picolinate) corresponds to S$_1$ at TDDFT level, while becomes S$_3$ in both BSE calculations. As a matter of fact, this is due to the inversion of the LUMO and LUMO+1
in the quasiparticle calculations.

By inspecting the frontier orbitals of Table~S2, we notice that in GW-BSE, the first singlet transition
display a MCLT with a larger metal contribution.
The case of Ir(ppy)$_3$ is simpler. Here the lowest singlets still present a mixed MLCT-CT character 
and are dominated by single transition
involving the same set of orbitals of FIrpic. The differences between TDDFT and BSE results are limited to the 
aforementioned modification in the adsorption frequencies. 
Note that lowest singlets are nearly degenerate, by virtue of the C$_3$ symmetry.
For the heteroleptic Ir(ppy)$_2$acac the S$_1$ and the S$_2$ are still dominated by a single transition
HOMO$\rightarrow$LUMO+1 and HOMO$\rightarrow$LUMO, respectively, in all the three approximations tested in this work. 
By consequence both these transitions are mixed MLCT and LC, with a limited contribution coming from the ancillary ligand.
On the contrary, more then one transition contribute to higher singlets, yielding to nontrivial differences between the various theoretical frameworks.
Singlet excications in Ir(ppz)$_3$ and Ir(thpy)$_3$ instead behave similarly. The only excitation which 
does not change among the three theoretical approaches is S$_1$, which is essentially a HOMO$\rightarrow$LUMO
transition (the orbital analysis proves a mixed MLCT and LC character of S$_1$ also for this complex), while for the the singlets immediately above the character of the transitions are comparable in the two BSE calculations but differ from TDDFT results.
In the last compound, Ir(btpy)$_2$acac, both S$_1$ and S$_2$ present a multiple transition character dominated by 
the contributions from HOMO$\rightarrow$LUMO and HOMO$\rightarrow$LUMO+1 (yielding to an admixture of MLCT and LC transitions at which the acac ligand contributes marginally) within all the three approaches, while 
the successive singlet excitations are sorted in a similar way for the BSE results but differ from the TDDFT ones.

The analysis of triplet excitations is somewhat more complex. 
For example, in the case of FIrpic the difference between the three approaches tested does not limit to the absolute value of the excitation energy, as mentioned at the beginning of 
the section. Indeed, the orbital components of the transitions are never the same comparing different methods.
Much easier is the case of Ir(ppy)$_3$ for which, similarly to singlet excitations, the triplets presents 
contributions coming from the same transitions both in TDDFT and in BSE based on both the starting points.
For Ir(ppy)$_2$acac the most significative difference is the switch of T$_5$, which becomes T$_1$ and T$_3$
in the BSE@G$_0$W$_0$ and BSE@G$_n$W$_n$ respectively.
The following two compounds addressed in the present work, Ir(ppz)$_3$ and Ir(thpy)$_3$, present an intricate behavior. 
Even the lowest energy triplet presents contributions originating from multiple transitions, and it is rather 
difficult to observe similarities between the different approaches adopted.
In the last compound, Ir(btpy)$_2$acac, the two lowest triplets present the same character 
in TDDFT and in BSE calculations, while the successive ones are consistent only comparing BSE calculations 
based on the two different GW flavors. Finally, we compared the G$_n$W$_n$-BSE excitations of FIrpic with
and without the Tamm-Dancoff Approximation (TDA), and we found that within TDA, the lowest-lying triplet
energies are higher by 0.02--0.08 eV.

\begin{figure}[htbp]
\begin{center}
  \includegraphics[width=8cm]{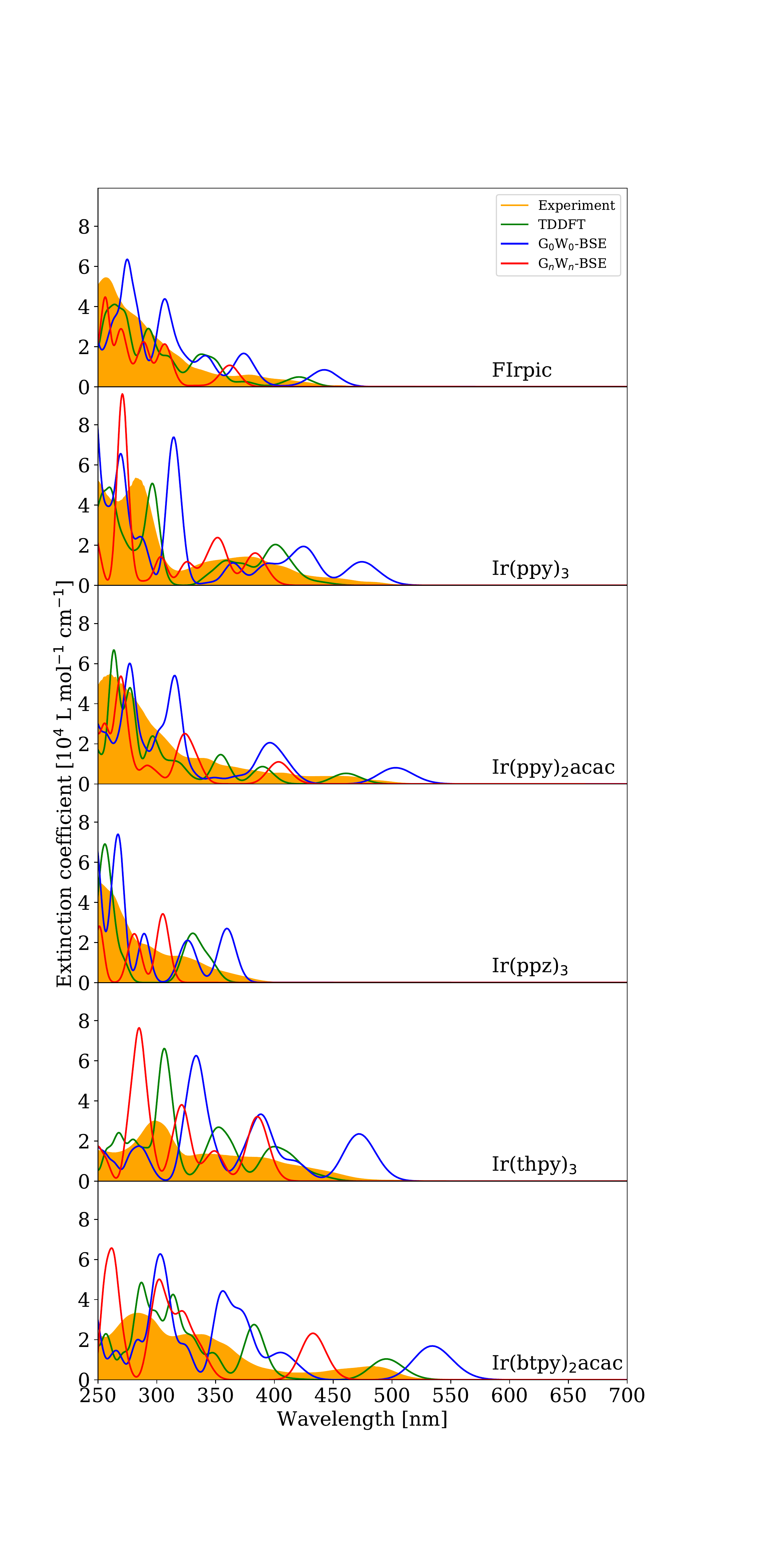}
  \caption{Extinction coefficient of the 6 complexes addressed in the present work.
  Calculations have been performed at the TDDFT-B3LYP level as well as from the BSE on top on G$_0$W$_0$ 
  and G$_n$W$_n$ results.}
  \label{fgr:complexesbse}
  \end{center}
\end{figure}

Figure \ref{fgr:complexesbse} reports the comparison of calculated extinction coefficients compared with measurments. 
As expected, the standard TDDFT-B3LYP level of theory produces theoretical spectra in fairly good agreement with experiments, providing adsorption edges close to the experimental ones and capturing all the main features presnt in the measurements.
The redshift of the excitation energies obtained from the BSE@G$_0$W$_0$ results yields to an overestimation of the wavelength at which these complexes starts to absorb light, as well as to a similar shift of the main features of the spectra.
On the contrary, the BSE@G$_n$W$_n$ extinction coefficients behave in the opposite way, presenting a blueshift of the adsorption edges and of the peaks of the spectra. 
To summarize, the TDDFT results provide a better agreement with experiments, with respect to the GW-BSE,
starting from B3LYP molecular orbitals.

\section{Functional dependence of the Ir(ppy)$_3$ properties}
\label{sec:irppy3}
To quantify the dependence of the electronic properties of Ir complexes on the choice of the exchange-correlation functional, we focussed on Ir(ppy)$_3$ and compared four DFT starting points, namely a pure local functional (BLYP), two standard hybrids with different percentage of exact exchange (B3LYP and BHLYP), and a Coulomb Attenuated functional (CAM-B3LYP). A similar study on the same complexes, only at the DFT/TDDFT level, can be found
in Ref.~\cite{Smith2016}.

\subsection{Quasiparticle energies}
\label{sec:irppy3_qp}
The HOMO-LUMO gaps of Ir(ppy)$_3$ calculated with the different approaches used in this work are tabled 
in Table \ref{tab:irppy3homolumo}.
As expected, at DFT level, the HOMO-LUMO gap increases with the increase of the percentage of 
exact exchange in $V_{xc}$, from a too much underestimated value of 2.1 eV with the pure BLYP functional to 
around 6 eV with the BHLYP and the CAM-B3LYP functionals.
Quasiparticle energies corrected by G$_0$W$_0$ display an opening of the HOMO-LUMO gap with 
respect to DFT results. Even if these results are closer to each other than DFT ones, as shown in benchmark
systems~\cite{Bruneval2013,Korbel2014,Jacquemin2015a}, they retain a significant dependence on the
DFT starting point.
On the contrary, when self consistency on eigenvalues is considered, the results are less sensitive to the approximation used for the DFT simulations, with HOMO-LUMO gaps ranging between 6.2 and 6.6~eV.

\begin{table}
  \caption{\label{tab:irppy3homolumo}HOMO-LUMO gap (eV) of Ir(ppy)$_3$ calculated as difference of 
  DFT-KS eigenvalues compared to G$_0$W$_0$ and G$_n$W$_n$ results.
  Calculations have been performed on top on a BLYP, B3LYP, BHLYP, and CAM-B3LYP DFT starting point.}
\begin{tabular}{|ll|lll|}
\hline
 && DFT & G$_0$W$_0$ & G$_n$W$_n$ \\
\hline
 BLYP      & Present              & 2.096 & 5.00 & 6.243 \\
 B3LYP     & Present              & 3.569 & 5.745 & 6.353 \\
           & Ref. \cite{Hay2002}  & 3.565    &          &          \\
           & Ref. \cite{Park2006} & 3.660    &          &          \\   
 BHLYP     & Present              & 5.860 & 6.541 & 6.654 \\
 CAM-B3LYP & Present              & 6.117 & 6.425 & 6.498 \\
 \hline
\end{tabular}
\end{table}

\begin{figure}[htbp]
\begin{center}
  \includegraphics[width=\textwidth]{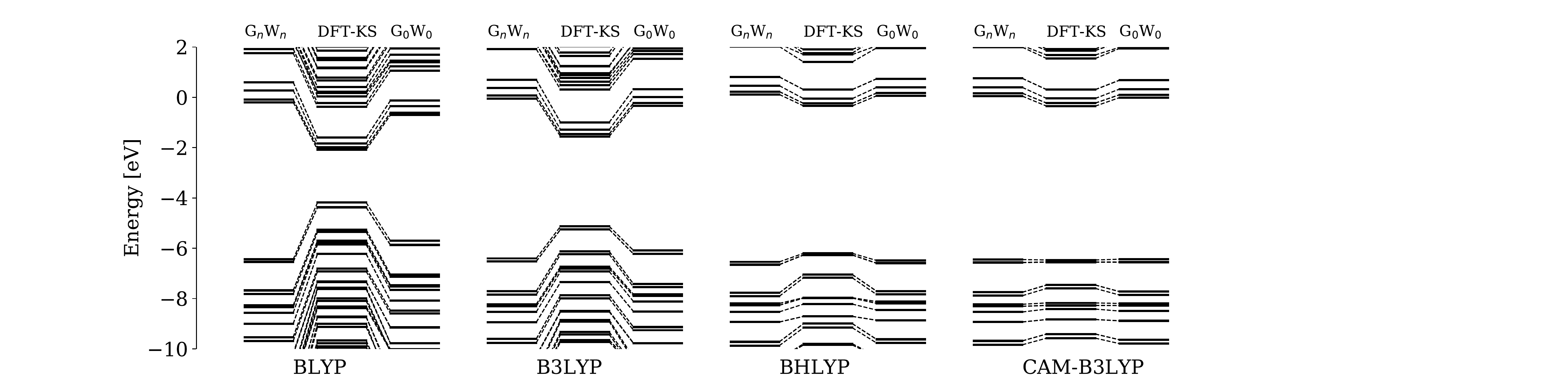}
  \caption{Quasiparticle energies of Ir(ppy)$_3$ obtained as DFT-KS eigenvalues compared to G$_0$W$_0$ 
  and G$_n$W$_n$ results.
  Calculations have been performed on top on a BLYP, B3LYP, BHLYP, and CAM-B3LYP DFT starting point.}
  \label{fgr:irppy3quasiparticle}
  \end{center}
\end{figure}

To visualize the effect of the different approximations tested in this section, we plot the quasiparticle 
energie in Fig. \ref{fgr:irppy3quasiparticle}.
Also in this case the opening of the HOMO-LUMO gap arises from the decrease of the energies 
of the occupied states and a comparable increase of the energies of the virtual states,
which is larger within G$_n$W$_n$ than with G$_0$W$_0$ for all the XC functionals tested.
In addition, the magnitude of the GW corrections is larger when a $V_{xc}$ with low percentage of exact exchange is choosen as 
starting point.

In Table~\ref{tab:irppy3ipea} we compare our calculated quasiparticle energies against experimental measurements of 
the Ionization Potential (IP) and Electron Affinities (EA), as well as calculations performed with the 
$\Delta$SCF method.
The Ionization Potential the results obtained from the $\Delta$SCF are rather close to both the 
G$_0$W$_0$ and G$_n$W$_n$ values, although they all overestimate the experimental results for the Ionization Potential.
A similar conclusion does not apply straightforwardly for the Electron Affinities, where in 
the particular case of the CAM-B3LYP functional the $\Delta$SCF method and the GW are in disagreement with the sign.
Additionally, the disagreement with the experimental value is much worse than for IP.

\begin{table}
  \caption{\label{tab:irppy3ipea} Ionization potential and electron affinity (eV) of Ir(ppy)$_3$ 
  calculated with the$\Delta$SCF approach compared to G$_0$W$_0$ and G$_n$W$_n$ results.
  Calculations have been performed on top on a BLYP, B3LYP, BHLYP, and CAM-B3LYP DFT starting point.}

  \begin{tabular}{|l|ll|}
\hline
 & I.P. & E.A.  \\ 
\hline 
 Experiment \cite{Yoshida2015} & 5.27 & 1.86\\
 Experiment \cite{DAndrade2005} & 5.10 & \\ 
\hline
 & \multicolumn{2}{c|}{BLYP} \\ 
 $\Delta$SCF & 5.887 & 0.610 \\ 
 G$_0$W$_0$ & 5.704 & 0.705 \\ 
 G$_n$W$_n$ & 6.441 & 0.198 \\ 
\hline
 & \multicolumn{2}{c|}{B3LYP} \\ 
 $\Delta$SCF & 6.187 & 0.451 \\ 
 G$_0$W$_0$ & 6.085 & 0.340 \\ 
 G$_n$W$_n$ & 6.403 & 0.050 \\ 
\hline
 & \multicolumn{2}{c|}{BHLYP} \\ 
 $\Delta$SCF & 6.127 & -0.142 \\ 
 G$_0$W$_0$ & 6.484 & -0.057 \\ 
 G$_n$W$_n$ & 6.540 & -0.114 \\ 
\hline
 & \multicolumn{2}{c|}{CAM-B3LYP} \\ 
 $\Delta$SCF & 6.294 & 0.078 \\ 
 G$_0$W$_0$ & 6.434 & 0.010 \\ 
 G$_n$W$_n$ & 6.451 & -0.048 \\ 
\hline
\end{tabular}

\end{table}

\subsection{Optical absorption}
\label{sec:irppy3_bse}
\begin{figure}[htbp]
\begin{center}
  \includegraphics[width=\textwidth]{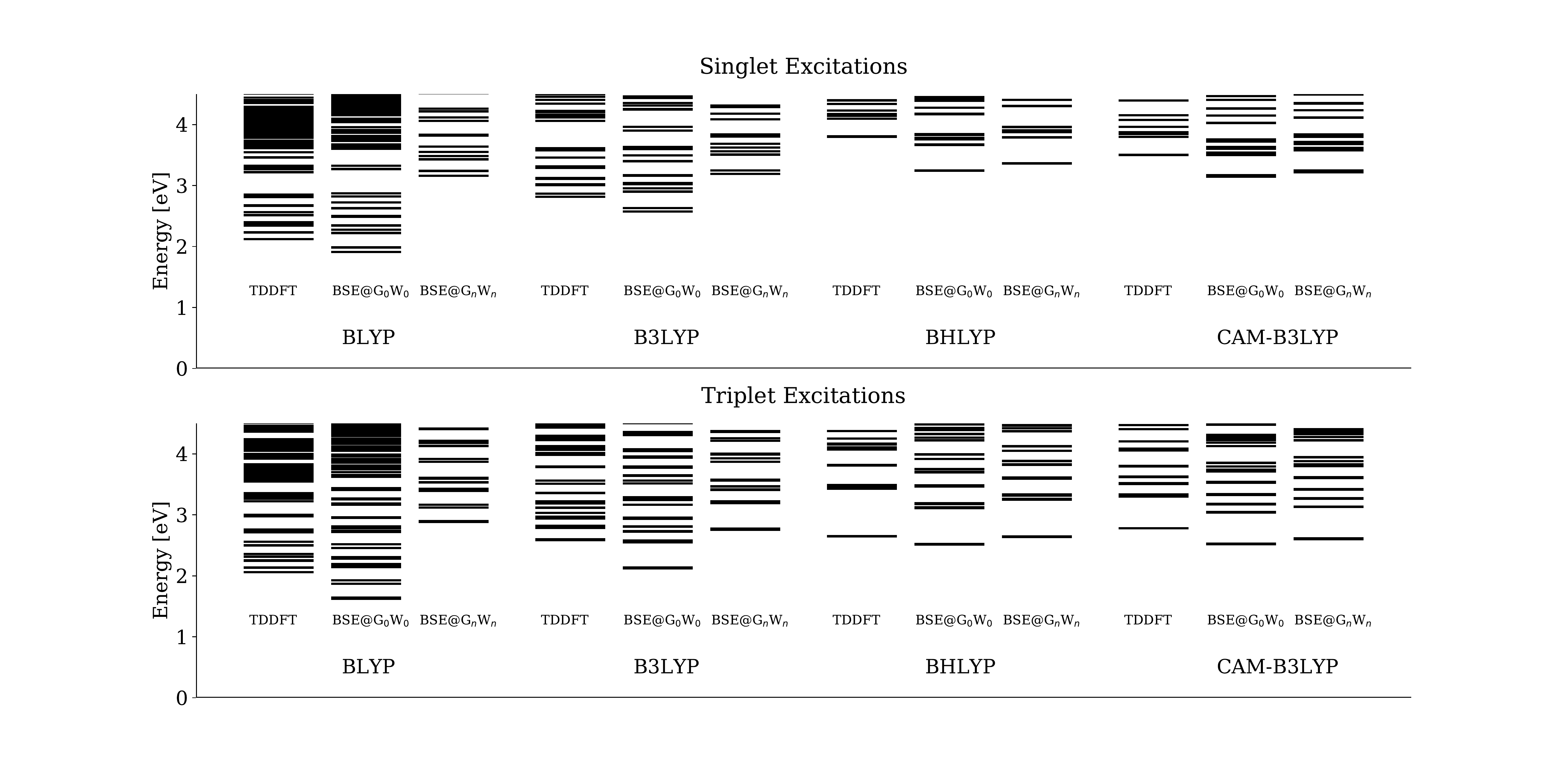}
  \caption{Lowest energy singlet and triplet excitation energies of Ir(ppy)$_3$ obtained at the 
  TDDFT level as well as from the BSE on top on G$_0$W$_0$ and G$_n$W$_n$ results.
  Calculations have been performed on top on a BLYP, B3LYP, BHLYP, and CAM-B3LYP DFT starting point.}
  \label{fgr:irppy3excitations}
  \end{center}
\end{figure}

Figure \ref{fgr:irppy3excitations} represents the behavior of the excitation energies for Ir(ppy)$_3$ singlets and triplets
calculated within TDDFT, BSE@G$_0$W$_0$ and BSE@G$_n$W$_n$, performed on top of DFT results obtained with four different $V_{xc}$.
At the TDDFT level the excitation energies of both singlets and triplets 
increase with increasing percentage of exact exchange in the $f_{xc}$ kernel.
Similarly to  quasiparticle energies, the excitation energies obtained from a BSE calculation performed on top of perturbative GW calculations, still retain a marked dependency on the DFT staring point and
they are always smaller than the respective TDDFT counterparts.
On the contrary, the results obtained from the BSE on top on G$_n$W$_n$ present a smaller dependence 
from the choice of the exchange and correlation functional, and only limited differences in the excitation
energies plotted in Fig. \ref{fgr:irppy3excitations} can be observed. 
These findings are overall consistent with previously published benchmarks.~\cite{Bruneval2015,Jacquemin2016a,Jacquemin2017b}.
Both singlets and triplets excitation energies computed at BSE@G$_n$W$_n$ level are always located at higher energies with respect to BSE@G$_0$W$_0$ ones. A general trend is not observed when comparing BSE and TDDFT results.
Indeed, while for BLYP and B3LYP functionals the BSE@G$_n$W$_n$ excitation energies are larger than 
the TDDFT ones, the opposite applies in the case of BHLYP and CAM-B3LYP. Energy differences among the three approaches reduces as the percentage of exact exchange in $V_{xc}$ increases, ranging from more than 1 eV in the BLYP case to about 0.1 eV when adopting the CAM-B3LYP functional.
 
The character and orbital contributions to the lowest excitations are reported in Table~S18-S25.
In particular, the lowest singlets present a comparable character for TDDFT, BSE@G$_0$W$_0$, and BSE@G$_n$W$_n$, when 
the BLYP and B3LYP functionals are used as starting point, while the comparison is less straightforward
for BHLYP and CAM-B3LYP. In these last cases, but for few transitions, the character of the transitions 
between TDDFT and BSE result present noticeable differences, while they involve mostly the same set of
orbitals in both the BSE calculations.

On the contrary, when the B3LYP functional was chosen as starting point, we did not observe differences between 
the TDDFT and BSE for the triplet transitions.
For the other V$_{XC}$ the character of the lowest triplet excitations remains the same between 
the two BSE methods but is different from TDDFT.
It is worth to notice how the lowest triplet present a character different from HOMO$\rightarrow$LUMO
only for TDDFT based on a BHLYP and CAM-B3LYP functionals.

\begin{figure}[htbp]
\begin{center}
  \includegraphics[width=8cm]{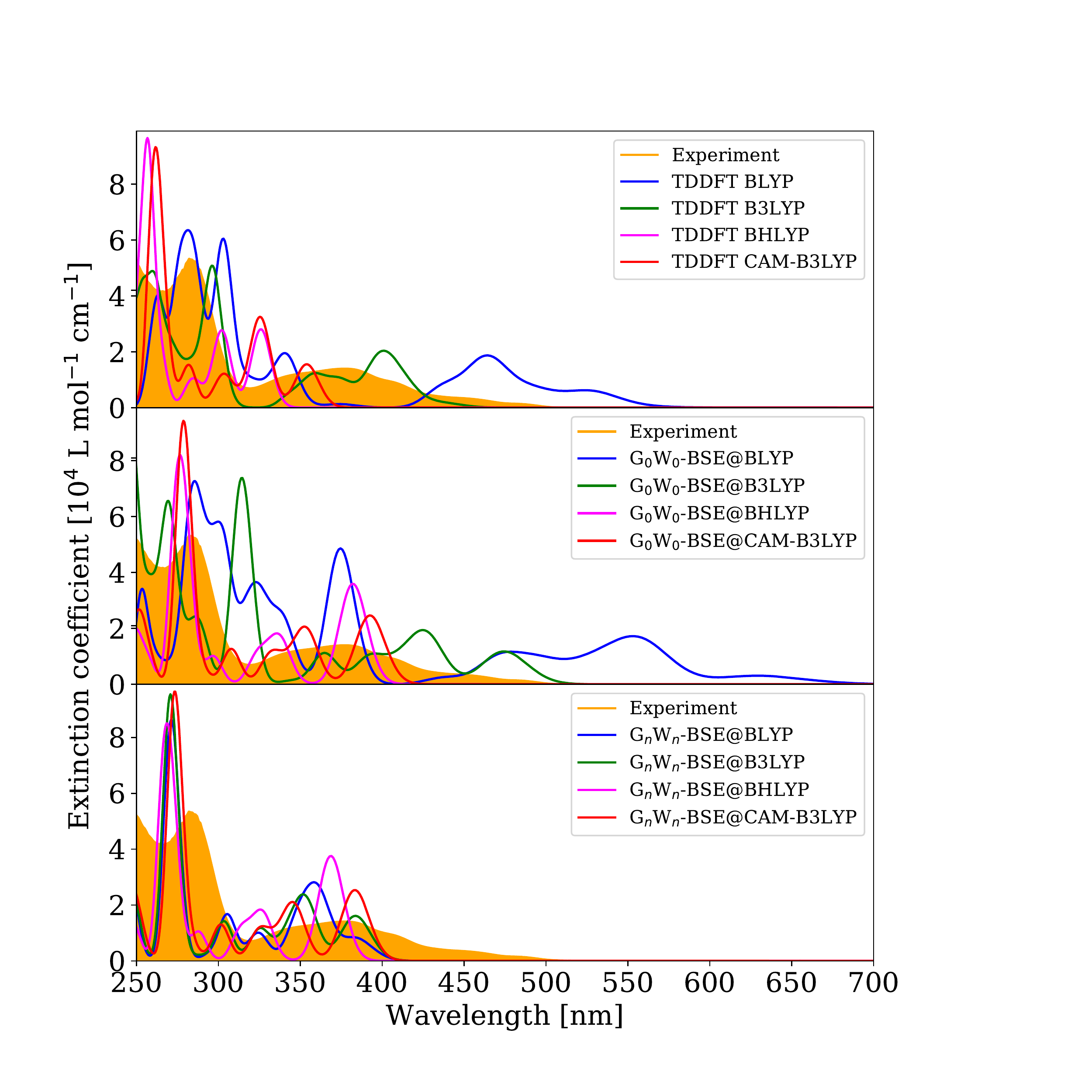}
  \caption{Extinction coefficient of Ir(ppy)$_3$ obtained at the  TDDFT level as well as from the 
  BSE on top on G$_0$W$_0$ and G$_n$W$_n$ results.
  Calculations have been performed on top on a BLYP, B3LYP, and CAM-B3LYP DFT starting point.}
  \label{fgr:irppy3bse}
  \end{center}
\end{figure}

We reporte in Figure \ref{fgr:irppy3bse} the comparison between the experimental extinction coefficient for 
Ir(ppy)$_3$ and the ones obtained from singlet excitations (we remark that to neglect of
spin-orbit coupling, as in the present study, causes transitions to triplet states to be forbidden).
In detail, TDDFT results show the well known trend, i.e. the B3LYP outcome reaches the best agreement with measurements. 
The pure BLYP functional produces red-shifted adsorption edges and peak position, while the remaining 
kernels behaves the other way round, yielding adsorption edges and peak positions at higher energies with respect to experimental measurements.
For any tested functional, the decrease of the excitation energies computed from the solution of BSE on top of perturbative GW 
calculations, causes the theoretical absorption spectra to move to higher wavelengths compared to the 
TDDFT ones. 
Therefore the main structures present in the BSE@G$_0$W$_0$ spectrum obtained on top of a B3LYP 
functional are significantly red-shifted with respect to TDDFT, deteriorating the agreement with experiment. 
To restore a reasonable agreement between theory and experiment it is required to introduce a larger fraction of exact exchange 
in V$_{XC}$. However, in the present case the main features of the BHLYP spectrum (similarly to the CAM-B3LYP one)
reside at too short wavelengths.
On the contrary, the results from a BSE calculation performed on top of G$_n$W$_n$ results are almost
independent on the underlying choice of V$_{xc}$, and only small differences can be observed in the position 
of the main structure of the spectra.
Anyway, when compared with experiment, all the tested staring points produces results which slightly underestimate
the wavelength of the main structures.

\section{Conclusions}
\label{sec:conclusions}
We calculated electronic and optical properties of six Ir(III) cyclometalated
complexes, widely used as light emitters in OLED devices, using the Many
Body Perturbation Theory method GW-BSE, both the one-shot G$_0$W$_0$ method,
and the eigenvalue self-consistent G$_n$W$_n$.
The quasiparticle levels (IP and EA) are well described by the $\Delta$SCF
and GW method, and the ionization potentials compare reasonably well to
available experimental data. The lack of experimental electron affinities
does not allow to draw a definite conclusion about the accuracy of the
calculated values.

The TDDFT@B3LYP calculated optical absorption spectra are in good agreement
with the experimental measurements carried out in this and previous works.
Starting from B3LYP wavefunctions, BSE@G$_0$W$_0$ results in a red-shift of
the spectra, whereas the BSE@G$_n$W$_n$ results in a blue-shift. We found
that by increasing the fraction of Hartree-Fock exchange in the
hybrid DFT calculations, the GW eigenvalue self-consistency is less
important, and the calculated BSE@G$_n$W$_n$ absorption spectrum is
less dependent from the DFT exchange-correlation, although blue-shifted
with respect to experiments.

Our calculations are among the largest GW-BSE calculations reported
in the literature~\cite{Deslippe2012,Govoni2015,Jacquemin2015a,Gui2018,Lettmann2019}
and they can provide realistic systems for testing the accuracy of current GW-BSE methods,
in order to explore the effects of different self-consistency schemes, and
of the underlying DFT exchange-correlation functional.

\section*{Supporting Information}
The Supporting Information is available free of charge on the
ACS Publications website at DOI: \url{10.1021/acs.jctc....}.

Frontier molecular orbitals of the six cyclometalated Ir(III) complexes;
single-particle (from DFT calculations) and quasi-particle
(from GW calculations) levels; full list of calculated excitation energies,
both at the TDDFT and GW-BSE level, with their main orbital transition
components.

\begin{acknowledgement}
The author thanks Fabien Bruneval for useful suggestions regarding the
use of the MolGW code. We acknowledge the CINECA and the Regione Lombardia
award under the LISA initiative 2016-2018 (project QUASOLED), for
the availability of high performance computing resources and support.
This project was partially funded by the GRO program of Samsung Advanced
Institute of Technology (SAIT).
\end{acknowledgement}

\bibliography{BSE_Ir_Complexes}

\end{document}